# Accurate and Generalizable Quantitative Scoring of Liver Steatosis from Ultrasound Images via Scalable Deep Learning

**Authors:** *Bowen Li[1]; *Dar-In Tai[2]; Ke Yan[1]; Yi-Cheng Chen[2]; Shiu-Feng Huang[3]; Tse-Hwa Hsu[2]; Wan-Ting Yu[2]; Jing Xiao[4]; Le Lu[1]; Adam P. Harrison[1]

1. PAII Inc., Bethesda MD, USA
2. Chang Gung Memorial Hospital, Linkou Medical Center, Taoyuan, Taiwan, ROC
3. Institute of Molecular and Genomic Medicine, National Health Research Institutes, Taipei, Taiwan, ROC
4. Ping An Insurance Group, Shenzhen, China

*equal contribution

Email of authors
Bowen Li: lbwdruid@gmail.com
Dar-In Tai: tai48978@cgmh.org.tw;
Ke Yan: yankethu@foxmail.com.
Yi-Cheng Chen: yicheng@cgmh.org.tw;
Cheng-Jen Chen: k85731@cgmh.org.tw;
Shiu-Feng Huang: sfhuang@nhri.org.tw;
Tse-Hwa Hsu: echohsuth45@cgmh.org.tw;
Wan-Ting Yu: ting800130@gmail.com
Jing Xiao: xiaojing661@pingan.com.cn
Lu Le: tiger.lelu@gmail.com
Adam P. Harrison: adam.p.harrison@gmail.com

## Keywords:

Ultrasound, Liver steatosis, Deep learning, Screening, Computer-aided diagnosis

## Corresponding Author:

Dr. Dar-In Tai. Division of Hepatology, Department of Gastroenterology and Hepatology, Chang Gung Memorial Hospital, Linkou Medical Center.
No 5, FuXing Street, Guishan Dist., Taoyuan 333, Taiwan, ROC.
Tel: 886-3-3281200 ext. 8107

Fax: 886-3-3272236

E-mail: tai48978@cgmh.org.tw

Dr. Adam P. Harrison, PAII Inc.
6720b Rockledge Drive, Bethesda, MD, USA, 20817.
E-mail: adam.p.harrison@gmail.com

## List of Abbreviations:

- Receiver Operating Characteristic (ROC); Deep Learning (DL); Control Attenuation Parameter (CAP); Magnetic Resonance Imaging Derived Proton Density Fat Fraction (MRI-PDFF); Nonalcoholic Fatty Liver Disease (NAFLD); Non-Alcoholic Steatohepatitis (NASH); Ultrasound (US); Chang Gung Memorial Hospital (CGMH); Big Data-Learning Group (***BD-L***); Big Data-Validation Group (***BD-V***); Histopathology-Validation Group (***HP-U***); Histopathology-Testing Group (***HP-T***); Tri-Machine Group (***TM***);   Left Liver Lobe (LLL); Right Liver Lobe (RLL); Liver Kidney Contrast (LKC); Subcostal (SC); Acoustic Radiation Force Impulse (ARFI); Hepatitis B Virus (HBV); Hepatitis C Virus (HCV); Non-Hepatitis B/Non-Hepatitis C (NBNC); Limit of Agreement (LOA); Repeatability Coefficient (RC); Area Under Curve Receiver Operating Characteristic (AUCROC)

## Financial Support:

This study is supported by the Maintenance Project of the Center for Artificial Intelligence in Medicine (Grant CLRPG3H0012, SMRPG3I0011) at Chang Gung Memorial Hospital and PAII Inc. (a US subsidiary company of Ping An Insurance Group).

# ABSTRACT

**Background & Aims:** Hepatic steatosis is a major cause of chronic liver disease. 2D ultrasound is the most widely used non-invasive tool for screening and monitoring, but associated diagnoses are highly subjective. We developed a scalable deep learning (DL) algorithm for quantitative scoring of liver steatosis from 2D ultrasound images.

**Approach & Results:** Using retrospectively collected multi-view ultrasound data from 3,310 patients, 19,513 studies, and 228,075 images, we trained a DL algorithm to diagnose steatosis stages (healthy, mild, moderate, or severe) from ultrasound diagnoses. Performance was validated on two multi-scanner unblinded and blinded (initially to DL developer) histology-proven cohorts (147 and 112 patients) with histopathology fatty cell percentage diagnoses, and a subset with FibroScan diagnoses. We also quantified reliability across scanners and viewpoints. Results were evaluated using Bland-Altman and receiver operating characteristic (ROC) analysis. The DL algorithm demonstrates repeatable measurements with a moderate number of images (3 for each viewpoint) and high agreement across 3 premium ultrasound scanners. High diagnostic performance was observed across all viewpoints: area under the curves of the ROC to classify $\geq$mild, $\geq$moderate, =severe steatosis grades were 0.85, 0.90, and 0.93, respectively. The DL algorithm outperformed or performed at least comparably to FibroScan with statistically significant improvements for all levels on the unblinded histology-proven cohort, and for =severe steatosis on the blinded histology-proven cohort.

**Conclusions:** The DL algorithm provides a reliable quantitative steatosis assessment across view and scanners on two multi-scanner cohorts. Diagnostic performance was high with comparable or better performance than FibroScan.

# INTRODUCTION

Liver steatosis, or fatty liver disease, is a major cause of chronic liver disease worldwide. It is estimated that non-alcoholic fatty liver disease (NAFLD) affects 20-30% of the global population[1–3] and is associated with increased risks of cardiovascular disease, type 2 diabetes, and metabolic risk factors[4]. Unfortunately, it is an under-treated and under-diagnosed disease[5]. For those patients with more aggressive non-alcoholic steatohepatitis (NASH), the risks of liver cirrhosis, liver failure, and hepatocellular carcinoma are higher[6–8]. Liver needle biopsy is the gold standard at present. However, biopsy's invasiveness severely limits its clinical applicability as a screening and assessment tool, and, at the same time, it is also prone to sampling error. Thus, with the growing prevalence of NAFLD and NASH, accurate, reliable, and accessible non-invasive screening tools are increasingly important to quantify liver steatosis and provide follow-up monitoring[4].

Such tools include magnetic resonance imaging with derived proton density fat fraction (MRI-PDFF), quantitative ultrasound (US), and 2D US diagnoses[9]. MRI-PDFF is prohibitively costly and time consuming for routine clinical care[5]. As for quantitative US, its most popular variant is FibroScan with its control attenuation parameter (CAP) scores[10]. FibroScan is more available than MRI-PDFF, but still requires dedicated equipment. On the other hand, 2D US exams have been widely used for the diagnosis of liver disease for 5 decades[9], which is partly driven by the prevalence of US equipment and its low cost. In clinical practice, 2D US is also the first-line screening modality for the detection of liver cancer[11], so steatosis can also be assessed from studies with a primary aim of liver tumor screening. Given these considerations, it is not surprising that 2D US is the most common tool for assessing liver steatosis[4]: a recent NAFLD epidemiology meta-analysis reveals that 90.6% of 392 studies in China used 2D US as the diagnostic modality of choice[3]. Unfortunately, US steatosis scores are considered a subjective diagnosis. A 2011 meta-analysis reports that the kappa statistics for inter- and intra-observer reliability show poor numbers[4]. A 2014 analysis[12], focusing on US steatosis assessment derived from routine clinical care, concludes that intra- and inter-observer agreements of binary assessment are only 51-68% and 39-40%, respectively, and it also notes that there is a lack of reported reliability measurements of categorical assessments. Both studies attribute the low reliability to different image acquisition practices across institutions and the subjective and variable nature of US image interpretation. More recently, Hong *et al.* have investigated the reliability of categorical US assessments of different features and reported only moderate inter-rater agreements (intra-class correlation coefficients of 0.54) for the overall steatosis impression[13].

A promising alternative is to apply machine learning algorithms, *e.g.*, deep learning (DL), on 2D US liver scans. The goal is to provide a quantitative measure of liver steatosis directly from 2D US images for clinical decision support. Efforts toward this end have been reported[14–22]. However, limitations in the analysis, *i.e.*, small training set sizes, single-scanner data[14–16,18–21] and only binary assessments [14–19], restrict the conclusions that can be drawn. Moreover, the reliability of these assessments, either across scanners or across different US views of the liver, have not been assessed. Relatedly, the specific number of images needed for a reliable diagnosis, and of which liver viewpoints, has also not been well articulated or characterized. This is crucial for an at-large adoption of any imaging-based diagnostic tool.

To address these limitations, we developed a scalable DL algorithm to quantitatively assess liver steatosis from 2D US, using a retrospectively mined big data cohort. Cross-scanner and cross-view reliability was measured, in addition to diagnostic performance against gold standard histopathological diagnoses on two clinical cohorts. Direct comparisons against FibroScan were also performed. The DL algorithm might serve as an effective tool for liver steatosis screening and monitoring.

# METHODS

## *Patient Cohorts and Image Collection*

Multiple patient cohorts were collected for our study, as shown in Table 1. The clinicopathological makeup of each dataset can be found in Supplementary Table 1a and 1b. All US images underwent the same automatic cropping and resampling preprocessing, which is described in the supplementary material. This study was approved by the Institutional Review Board (IRB) of the Chang Gung Medical Foundation (CGMH IRB No.: 201801283B0).

**Table 1.** Overview of development and testing datasets

| Stage | Name | Purpose | Labels | Patients | Studies | Images |
|---|---|---|---|---|---|---|
| DL Learning | **BD-L** | Big data, to train the neural network | 2D US Dx | 2,899 | 17,149 | 200,654 |
| DL Validation | **BD-V** | Big data, to tune model performance | 2D US Dx | 411 | 2,364 | 27,421 |
| Testing | **HP-U** | Histopathology-proven group, to (a) measure the trend between DL predictions and histology (b) measure reliability across 2D US liver viewpoints | Histology | 147 | 147 | 1,647 |
| Testing | **TM** | Tri-machine data US Dx group, to (a) measure reliability across 2D US liver viewpoints and (b) measure reliability across scanners | -- | 246 | 733 | 9,215 |
| Testing | **HP-T\*** | Histology proven group to measure the trend between DL predictions and histology | Histology, FibroScan | 112 | 112 | 1,996 |

\* Labels blind to DL researchers during course of algorithmic development

**Big data groups (*BD-L* & *BD-V*)**

We retrospectively collected a big-data dataset from the picture archiving and communication system of Chang Gung Memorial Hospital (CGMH), a major hospital in Taiwan (over 4 million outpatient visits/year). All patients who received elastography (a quantitative US technology), specifically acoustic radiation force impulse (ARFI) imaging and FibroScan, between 01/03/2011 and 09/28/2018 represented the index patients. From the index patients, we extracted all 2D US studies that were acquired within the same 2011-2018 period, resulting in multiple studies per patient. 70% of the patients, and their corresponding US studies, were randomly selected as training data (***BD-L***), totaling 2,899 patients and 200,654 images. We used another 10% of the patients for validating and tuning our DL algorithm (***BD-V***), totaling 411 patients and 27,421 images. The remaining 20% of the patients were not used as part of this study. *In addition, any patients also found in the other datasets were also moved to this excluded cohort.* The collected images were generated from 13 known scanners, which are listed in Supplementary Table 2. Each US study is accompanied by a 2D US diagnosis of steatosis severity from visual assessment, generated through the course of routine clinical care. These labels are used to train the DL algorithm.

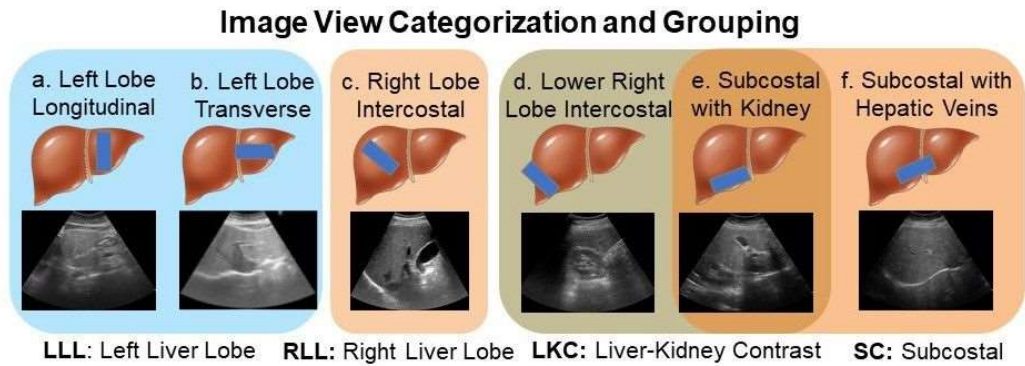

**Figure 1**. Image view categorization and grouping. Six US image viewpoints are used in this study: a. left lobe longitudinal, b. left lobe transverse, c. right lobe intercostal, d. lower right lobe intercostal (depicting liver/kidney contrast), e. subcostal depicting liver/kidney contrast, and f. subcostal without hepatic veins. These views are further categorized into four groups: left liver lobe (LLL, a and b), right liver lobe (RLL, c), liver/kidney contrast (LKC, d and e), and subcostal (SC, e and f). Liver cartoons adapted from the DataBase Center for Life Science (https://commons.wikimedia.org/wiki/File:201405_liver.png), licensed under the Creative Commons Attribution 4.0 International.

**Unblinded Test Group (*HP-U*)**

We used a collection and selection protocol reported in prior work[23–26]. Specifically, since 2011 it has been CGMH policy to obtain elastography measurements for all patients undergoing a liver biopsy, *i.e.*, ARFI and FibroScan, once the latter became available. The ***HP-U*** dataset (n=147) consists of patients with FibroScan diagnoses between the dates 11/27/2014 to 09/26/2019. Identical to a prior study[23], we included patients with chronic hepatitis B virus (HBV), chronic hepatitis C virus (HCV), and non-hepatitis B/C virus (NBNC) liver diseases. We excluded those with decompensated liver disease, toxic hepatitis, alcoholism, or autoimmune liver diseases. After exclusion of these etiologies, the majority of the NBNC group are NAFLD patients (90.3% in ***HP-U*** group, supplementary Table 1b). Histopathological analysis was performed by the same clinician (S.F.H.) and any retrospectively collected histopathological analysis not originally performed by S.F.H. was redone, which avoids potentially large inter-rater variability[27,28] from confounding the gold standard. Following Kleiner *et al.*[29], the histology diagnoses were graded into normal (<5%), mild (≥5% & <33%), moderate (≥33% & <66%) and severe (≥66%) based on the liver fat cell fraction.

Because we are interested in US analysis, additional selection criteria were also applied. A patient must have a 2D US study that was: (1) acquired within 3 months of the biopsy; and (2) acquired with one of the Siemens Acuson S2000, Philips IU22, or Toshiba Aplio 300 scanners. We also excluded patients with tumors > 3cm and with multiple cysts. Finally, US studies must have >=10 images of the viewpoints depicted in Figure 1. If more than one US study qualified, we randomly selected one. All labels in *HP-U* were unblinded to the DL researchers of this work, but the data was treated as a test set, meaning it was only analyzed after development was complete.

**Tri-machine group (*TM*)**

We prospectively collected this cohort (n=246) from patients that had both an US and a Fibroscan study ordered and if D.I.T., Y.C.C., T.H.H. or C.J.C. were conducting the image study. With the agreement of these patients, they were scanned by Siemens Acuson S2000, Philips IU22, Toshiba Aplio 300 scanners on the same day. Studies were collected over a period from 08/30/2018 to 08/27/2019, and we only included patients diagnosed with the HBV, HCV, and NBNC criteria used in *HP-U*. The *TM* cohort allows for an assessment of agreement across scanners.

**Blind Testing Group (*HP-T*)**

Finally, we included *HP-T* (n=112), a clinical testing dataset whose labels were blind to the DL researchers involved in this project during development of the algorithm. *HP-T* was collected from patients that had received a liver biopsy between 03/18/2011 to 05/05/2015 and 09/01/2018 to 01/29/2021. Associated FibroScan diagnoses are only available for the later subset of patients. Also, unlike *HP-U*, US studies were not restricted to only the Siemens, Philips, and Toshiba scanners of *TM* and *HP-U*. In particular, 18 studies were acquired with the Aloka SSD 5500 or ATL: HDI 5000 scanners, which are not currently considered premium scanners.

## *Image Selection*

We are interested in investigating performance and reliability across viewpoints. Thus, we only included US images from the viewpoints shown in Figure 1, which can be categorized into four *view groups*: left liver lobe (LLL), right liver lobe (RLL), liver/kidney contrast (LKC), and subcostal (SC). For *HP-U, HP-T*, and *BWC*, we only included studies that had >=10 images of any of the studied viewpoints. *BD-L* and *BD-V* were automatically filtered with an algorithm explained in the Supplementary.

## *Training Steatosis Assessment DL Algorithm*

Figure 2 illustrates the algorithmic workflow of our DL algorithm. Using the images from **BD-L**, we trained a multi-class deep ResNet18[30] DL classifier using the 2D US diagnoses, which are ordinal labels ranging from 0 to 3 corresponding to None; Mild; Moderate; and Severe steatosis[31]. We treat each image independently in training and follow the well-known binary decomposition approach to ordinal classification of Frank and Hall[32]. After training, a simple transformation produces a *continuous* score[33] for each image that ranges from 0 to 1, with higher scores corresponding to more severe steatosis. As Figure 2 indicates, during inference, we take the mean of image-wise scores within and across each view group. The view group scores are further averaged to produce an "All View Groups" score. If one or several view groups are missing, the "All View Groups" score is calculated with what view groups are available. More details on the training strategy can be found in the Supplementary and a listing of hyper-parameters are given in Supplementary Table 5.

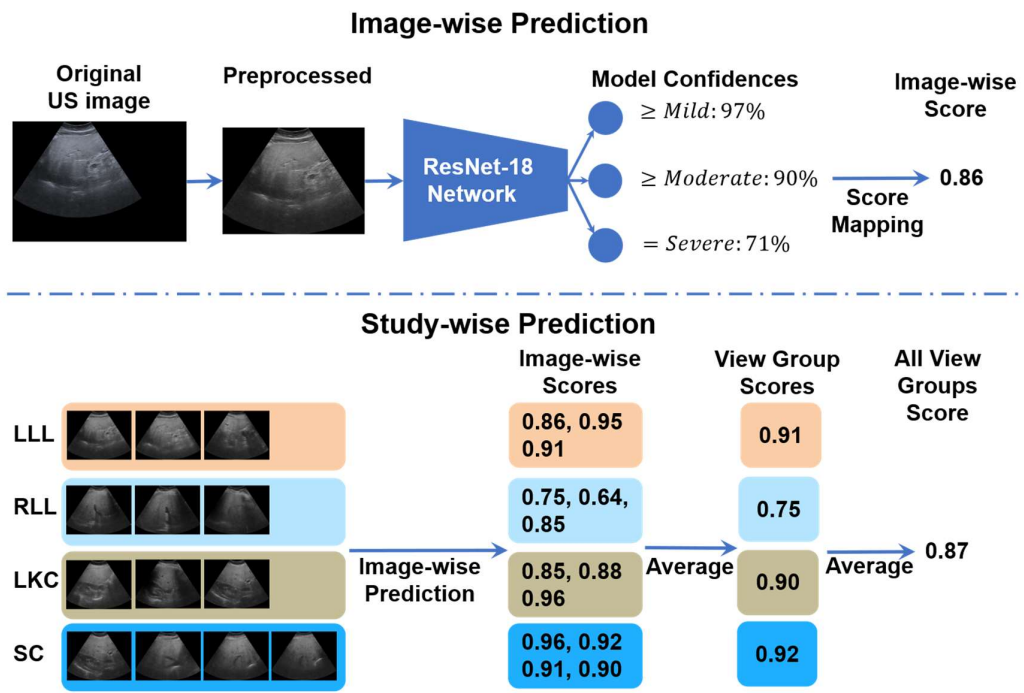

**Figure 2.** Algorithmic Workflow. Images are first comprehensively pre-processed to remove regions outside the US beam. A DL neural network, called ResNet-18, is trained on individual US images in **BD-L**. The model predicts confidences in three binary cutoffs: ≥mild, ≥moderate, or =severe steatosis. The confidences are mapped to a continuous *image-wise* score in the range of [0,1]. View-group scores are produced by averaging each image within the group. An "All View Groups" score is produced by averaging all available view group scores. In the figure's example, the gold standard histopathology diagnosis is a fatty cell percentage of 90%.

## *Statistical Analysis*

A listing of all experiments can be found in Supplementary Table 3. We evaluated the DL algorithm's reliability and diagnostic performance. We judge *p* values <0.05 as significant and corrected for multiple comparisons using the Holm-Bonferroni procedure [34].

**Reliability Studies**

**Experiment 1**: We used **TM** and **HP-U** to assess how many images are needed per view group to achieve repeatability. Note, for the **TM** dataset we randomly selected only one US study for each patient to avoid sampling the same patient more than once. As advocated by Bland and Altman[35,36], we graphed the within-subject standard deviations across different view-group steatosis scores and measured the repeatability coefficient (RC) [36]. The difference between two repeated measurements should be within the RC value for 95% of the US studies. However, because the within-subject standard deviation is not uniform across our data (typically greater variability in repeated measurements at moderate steatosis levels), we regressed a non-uniform RC and used the worst-case RC value as a summary statistic, with 95% confidence intervals computed using percentile bootstrap (1000 bootstrap samples)[37]. More details on this calculation can be found in the Supplementary.

**Experiment 2**: We also evaluated agreement across scanners using **TM**: we conducted a Bland-Altman analysis[35] on the difference values across view-group scores and calculated the limits of agreement (LOAs[36]), which are the limits by which 95% of the disagreements fall under[36]. Like Experiment 1, variability tended to be higher at moderate levels of severity, so we regressed non-uniform LOAs and used the maximum upper LOAs and minimum lower LOAs as summary statistics. More details can be found in the Supplementary. We also calculated the percentage agreement, using the cutoff levels determined in Experiment 3 below.

**Diagnostic Testing**
**Experiment 3a**: We validated our DL model's diagnostic performance using the images and histopathological labels of **HP-U**. Although the **HP-U** labels were not blinded to the DL researchers during algorithmic development, it was treated like a test set, *i.e.,* evaluation was only performed once model development was complete. Histopathological diagnoses were separated into four ordinal labels[27]. For three separations of fatty percentages, *i.e.,* grade >=5%, grade >=33%, and grade >=66%, we used receiver operating characteristic (ROC) curve analysis and measured the area under the curve of the ROC (AUCROC). Trend tests between the DL assessment and histopathological grades were conducted using the non-parametric

Jonckheere-Terpstra test[38]. The 95% confidence intervals of all AUCROCs were calculated based on DeLong's non-parametric test[39]. When needed, we determined cutoff values using the values that maximized the Youden index[40]. **Experiment 3b:** We compared the performance of FibroScan with that of the DL assessment, and statistical significance of any differences in AUCROCs were assessed using the StAR[41] implementation of DeLong's non-parametric test[39].

**Experiment 4a**: Finally, we tested our DL assessment on *HP-T*, whose histopathological labels were blind to the DL researchers during model development. **Experiment 4b**: For patients with FibroScan diagnoses, we also compared its performance with that of the DL assessment. Experiments 4a-b used the same statistical analyses as Experiments 3a-b.

# RESULTS

## *Repeatable Measurements with a Moderate Number of Images*

We first determined how many images are needed for each view group to reach a repeatable measurement (experiment 1 in Supplementary Table 3) using ***TM*** (one random study per patient) (n = 246) and ***HP-U*** (n = 147). The results when using three images per view group can be found in Table 2, and the complete set of results from one-to-four images can be found in Supplementary Table 6. As can be seen, depending on the view group, the max RC value is equal to or less than 30% of the DL assessment scale, which ranges from 0 to 1. In the worst case, 95% of the differences between repeated measurements should be within this max RC value. As the repeatability graph of Figure 3a demonstrates, this max RC value occurs at moderate levels of steatosis severity, which have a wide tolerance (*e.g.,* moderate fatty cell content is commonly determined as being anything between 33 to 66%). The RC values for mild and severe are much lower, suggesting that a 30% RC value is highly conservative for these ranges of severity. The repeatability graphs for all other view groups can be found in Supplementary Figure 1. Together, these data demonstrate the DL assessment can attain repeatable measurements with a moderate number of images for each view group. *For the remainder of evaluations, we will only include assessments that have a minimum of three images for each view group.*

**Table 2.** Reliability studies. On the left, the max repeatability coefficient (RC) is tabulated when using *three* images for each view group (***TM*** and ***HP-U*** datasets). On the right, the bias, worst-case limits of agreement (LOA), and % agreement are tabulated across different view groups for the ***TM*** dataset. Here, the results for all scanner pairs are combined. Parentheses enclose bootstrapped 95% confidence intervals.

|  | **Repeatability Study** | | **Cross-Scanner Agreement Study** | | | | |
|---|---|---|---|---|---|---|---|
| View | $N$ | $\max RC$ | $N$ | *Bias* | min *lower LOA* | max *upper LOA* | Agreement |
| LLL | 342 | 0.27 (0.24, 0.29) | 237 | 0.00 (-0.01, 0.01) | -0.37 (-0.32, -0.42) | 0.37 (0.32, 0.42) | 92% |
| RLL | 370 | 0.21 (0.19, 0.23) | 232 | 0.00 (-0.01, 0.01) | -0.37 (-0.33, -0.42) | 0.37 (0.33, 0.42) | 92% |
| LKC | 267 | 0.30 (0.27, 0.34) | 183 | 0.00 (-0.01, 0.01) | -0.35 (-0.29, -0.41) | 0.35 (0.29, 0.41) | 93% |
| SC | 297 | 0.26 (0.24, 0.29) | 182 | 0.00 (-0.01, 0.01) | -0.36 (-0.31, -0.42) | 0.36 (0.31, 0.42) | 94% |
| All View Groups | -- | -- | 237 | 0.00 (-0.01, 0.01) | -0.25 (-0.21, -0.28) | 0.24 (0.21, 0.28) | 94% |

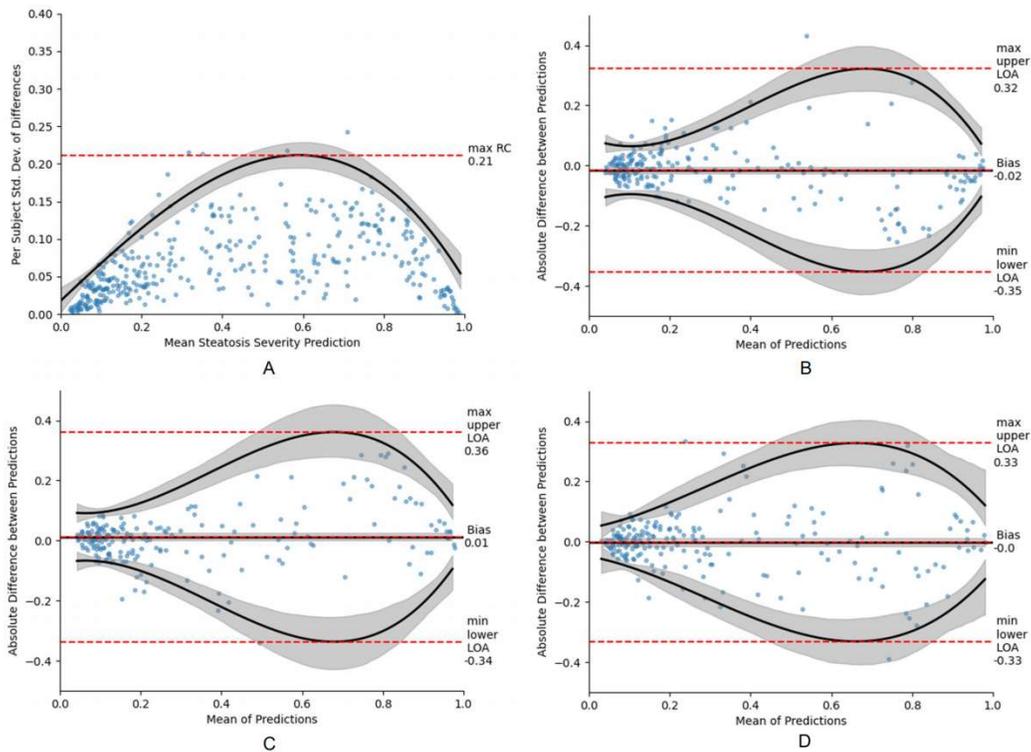

**Figure 3:** (A) shows a repeatability coefficient (RC) plot for RLL when using three images, while (B) to (D) represents cross-scanner Bland-Altman plots for Siemens-Toshiba, Toshiba-Philips, and Philips-Siemens, respectively. Cross-scanner plots are depicted for 'All View Groups' when using >= three images per view group. Grey-shaded areas indicate 95% confidence intervals.

### *Agreement Across Scanners is High*

We used the *TM* dataset (n = 246) to measure agreement across measurements taken on the same day but with three different scanners (experiment 2 in Supplementary Table 3). As the Bland-Altman plots for "All View Groups" of Figure 3b-d demonstrate, the LOAs are worse at moderate levels of severity, but they are much tighter at milder and more severe levels. This mirrors the repeatability results. Table 3 presents the Bland-Altman summary statistics for all view groups, demonstrating that the cross-scanner agreement is roughly equivalent for all view groups. For brevity we combined the results across all scanner pairs together. As can be seen, the agreement across scanners is high. The bias for "All View Groups" is close to zero and the worst case LOAs are around 35% for every view group. The LOAs for "All View Groups" fall to 25%, suggesting that examining all view groups together can increase reliability. The percentage agreement numbers are all 92% or higher, further underscoring the high agreement across the tested scanners.

### *High Diagnostic Performance on Clinical Test Sets*

With the reliability studies completed, we validated the DL score against a histopathological gold standard for diagnosing mild-to-severe (>=5%), moderate-to-severe (>=33%) and severe (>=66%) fatty percentages. The results on HP-U and HP-T, experiments 3 and 4 in Supplementary Table 3, respectively, are presented in Table 3. Focusing primarily on the HP-T results, the "Complete 4 view groups study" of Table 3 only selects studies that have three or more images for every view group. This allows an "apples-to-apples" comparison across view groups. As can be seen, the performance is comparable across view groups, suggesting that each view group provides similar diagnostic value, with all providing AUCROCs >=0.84. The "Individual view groups study", on the other hand, presents results when examining each view group individually, meaning only the view group in question requires three or more images. This allows for a larger sample size. In this setting "All View Groups" denotes the mean score across all view groups with three or more images, which can comprise different view groups from study to study. The ROC curves for "All View Groups" for the individual view group study can be found in Figure 4a. As can be seen in Table 3, the results for the individual view groups are broadly like those of the complete view groups study, with AUCROCs >=0.81. Comparing the results of HP-U to HP-T, the AUCROCs are generally similar, reinforcing the above results. However, the HP-U results are generally better, especially for diagnosing mild-to-severe steatosis. This is due to HP-T including non-premium Aloka and ATL HDI scanners in its cohort, which make it harder to accurately assess steatosis. Indeed, as Figure 4b indicates, when only selecting for the Siemens, Philips, and Toshiba premium scanners the AUCROC scores for HP-T are much improved, indicating that scanner choice does make an impact. Finally, the "FibroScan comparison study" compares the performance of FibroScan directly with the DL assessment. As can be seen, the DL assessment AUCROC values were better than FibroScan, and statistical significance was reached for all levels on the HP-U dataset, and for =severe on the HP-T dataset.

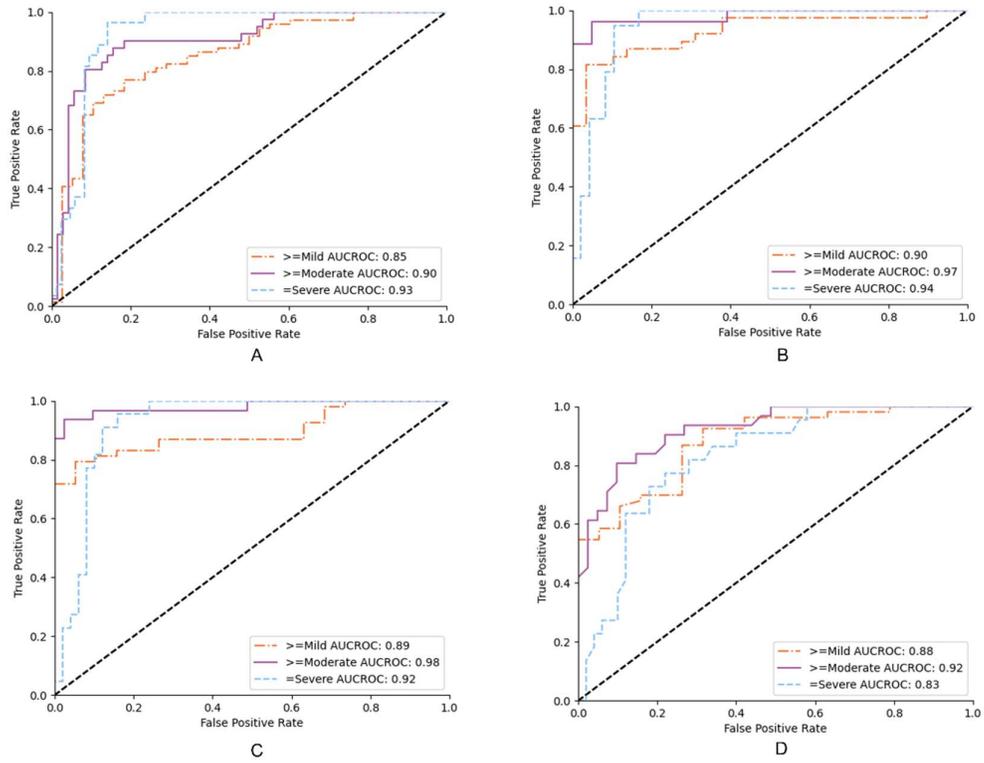

**Figure 4:** (A) and (B) depict ROC curves of the DL model for diagnosing hepatic steatosis grades on *HP-T*, when using all scanners and only the Siemens/Toshiba/Philips premium scanners, respectively. (C) and (D) only select for *HP-T* studies with FibroScan diagnoses, corresponding to the performance of the DL algorithm and FibroScan, respectively. All ROC curves are measuring against a histopathological gold standard.

**Table 3.** ROC analysis on the *HP-U* and *HP-T* cohorts for diagnosing steatosis grades. "Complete 4 view groups study" only selects studies where *every* view group is qualifying (3 or more images), whereas "Individual view group study" examines the performance of each qualifying view group individually. "FibroScan comparison study" only selects for studies with corresponding FibroScan diagnoses. Numbers in parentheses are 95% confidence intervals. All trends between the DL/FibroScan score and the histopathology grades were significant ($p<0.001$). "Acc" is the classification accuracy when the threshold values calculated by optimizing the Youden index[46] are applied.

| View | \| | N | AUC: ≧ 5% | AUC: ≧ 33% | AUC: ≧ 66% | Acc | \| | N | AUC: ≧ 5% | AUC: ≧ 33% | AUC: ≧ 66% | Acc |
|---|---|---|---|---|---|---|---|---|---|---|---|---|
| | | | | HP-U | | | | | | HP-T | | |
| • *Complete 4 view groups study* * | | | | | | | | | | | | |
| LLL | | 41 | 0.98 (0.93, 1.00) | 0.95 (0.89, 1.00) | 0.94 (0.87, 1.00) | 83% | | 51 | 0.90 (0.82, 0.98) | 0.92 (0.82, 1.00) | 0.90 (0.81, 0.99) | 88% |
| RLL | | 41 | 0.96 (0.90, 1.00) | 0.95 (0.89, 1.00) | 0.89 (0.79, 0.99) | 85% | | 51 | 0.86 (0.76, 0.96) | 0.93 (0.85, 1.00) | 0.92 (0.85, 1.00) | 96% |
| LKC | | 41 | 0.96 (0.90, 1.00) | 0.95 (0.89, 1.00) | 0.93 (0.84, 1.00) | 83% | | 51 | 0.84 (0.73, 0.95) | 0.95 (0.90, 1.00) | 0.88 (0.79, 0.97) | 90% |
| SC | | 41 | 0.96 (0.90, 1.00) | 0.92 (0.84, 1.00) | 0.89 (0.79, 0.99) | 83% | | 51 | 0.88 (0.79, 0.97) | 0.93 (0.86, 1.00) | 0.88 (0.77, 0.99) | 88% |
| All View Groups | | 41 | 0.96 (0.90, 1.00) | 0.94 (0.88, 1.00) | 0.92 (0.83, 1.00) | 83% | | 51 | 0.89 (0.80, 0.98) | 0.95 (0.88, 1.00) | 0.91 (0.83, 0.99) | 94% |
| • *Individual view group study* * | | | | | | | | | | | | |
| LLL | | 103 | 0.95 (0.90, 0.99) | 0.93 (0.87, 0.98) | 0.91 (0.86, 0.97) | 80% | | 96 | 0.84 (0.76, 0.93) | 0.92 (0.85, 0.99) | 0.93 (0.88, 0.98) | 90% |
| RLL | | 138 | 0.94 (0.91, 0.98) | 0.91 (0.86, 0.98) | 0.85 (0.78, 0.92) | 83% | | 109 | 0.83 (0.75, 0.91) | 0.90 (0.83, 0.96) | 0.92 (0.87, 0.97) | 92% |
| LKC | | 88 | 0.96 (0.92, 1.00) | 0.92 (0.86, 0.98) | 0.84 (0.76, 0.92) | 80% | | 71 | 0.81 (0.69, 0.93) | 0.93 (0.87, 0.99) | 0.89 (0.81, 0.96) | 90% |
| SC | | 117 | 0.93 (0.89, 0.98) | 0.91 (0.85, 0.96) | 0.86 (0.79, 0.92) | 79% | | 90 | 0.86 (0.77, 0.94) | 0.89 (0.82, 0.96) | 0.90 (0.83, 0.97) | 87% |
| All View Groups | | 147 | 0.95 (0.91, 0.98) | 0.92 (0.88, 0.96) | 0.87 (0.81, 0.92) | 76% | | 112 | 0.85 (0.77, 0.93) | 0.90 (0.84, 0.96) | 0.93 (0.88, 0.98) | 90% |
| • *FibroScan comparison study* *† | | | | | | | | | | | | |
| All View Groups | | 147 | 0.95‡ (0.92, 0.98) | 0.95‡ (0.92, 0.98) | 0.92‡ (0.88, 0.97) | 77% | | 80 | 0.89 (0.82, 0.96) | 0.98 (0.95, 1.00) | 0.92‡ (0.86, 0.98) | 93% |
| FibroScan | | 147 | 0.88 (0.81, 0.95) | 0.88 (0.81, 0.95) | 0.80 (0.73, 0.87) | 62% | | 80 | 0.88 (0.80, 0.96) | 0.92 (0.86, 0.98) | 0.83 (0.73, 0.92) | 68% |

\* Filtered with a minimum of 3 images for each view group
† Selecting only studies with associated FibroScan CAP scores
‡ AUCROC significantly better than FibroScan

# DISCUSSION

## *A Versatile and Reliable Steatosis Assessment*

Incorporating different 2D US scanner models and brands, different liver viewpoints, and prospectively and retrospectively collected images, we demonstrated that a DL-based assessment can provide quantitative and reliable hepatic steatosis scores. Unlike ATI[42], FibroScan, or some other reported DL solutions[17,20,21] (Supplementary Table 7), our algorithm accepts images taken from both hepatic lobes rather than a selected area of interest in a specific location. We categorize 2D US images into 6 major viewpoints (Figure 1), which we further group into 4 view groups (LLL, RLL, RKC and SC). We found that for each view group three images were enough to reach acceptable max RC values of 21-30% (Table 2, Figure 3a), where the best and worst max RC values corresponded to the RLL and LKC view group, respectively. The relatively poorer repeatability of the LKC view group is likely due to the heterogenous makeup of viewpoints, as it comprises both sub-costal and right lobe intercostal images (see Figure 1). To put these repeatability ranges in context, the RC for gold-standard histopathology fatty percentage assessment has been reported to be 38%[27] with poor intraclass correlation (ICC) agreements of 0.57[28]. Considering our DL model's worst max RC value is 30%, and that RC values tended to be much better than 30% at milder or more severe levels of steatosis (see Figure 3a and Supplementary Figure 1), these repeatability measures compare well. It is also encouraging that the DL algorithm is more repeatable at milder and severe steatosis levels since such patients should indeed be less ambiguous to categorize. In addition, we also demonstrate good cross-scanner agreement. Bland-Altman analysis suggests bias across scanners is near zero with acceptable LOAs (35-37%) for individual view groups, with the LOAs falling to 25% when using "All View Groups" (Table 2; Figure 3b; Supplementary Figure 1). When using categorical labels, these numbers correspond to agreements of 90 to 96%. An encouraging sign for generalizability is that cross scanner agreement is high for the Siemens: S2000 scanner, despite it being poorly represented in *BD-L* (Supplementary Table 2).

In terms of diagnostic performance, we validate on two different histology-proven cohorts (*HP-U* and *HP-T*). The "Complete view group study" of Table 3 indicates that diagnostic performance remains stable across view groups, with comparable and high AUCROC values (>=0.84). These results are reinforced by the "Individual view group study", which allowed each view group to be investigated individually, with a corresponding larger sample size. Again, the DL model posted good AUCROC values (>=0.81) across view groups and histopathology grades. As highlighted in the results, the AUCROCs in *HP-T* tended to be lower than in *HP-U*, which was most pronounced for diagnosing >5% fatty percentage. This is due to 18 earlier studies that were included in **HP-T** (Supplementary Table 2), which produced relatively lower quality images than the more recent Philips, Siemens, and Toshiba premium scanners. Indeed, as Figure 3b shows, excluding the older scanners raised the **HP-T** AUCROC scores considerably so that they matched **HP-U**'s (from 0.85 to 0.90 AUCROC for >=5% fatty percentage). Consequently, even though performance seems to be stable across the tested premium scanners, the DL algorithm can be sensitive to scanner quality, which is a matter requiring further investigation.

The DL algorithm demonstrated good quantitative performance despite being trained on subjective 2D US categorical labels. We speculate this is due to the exceptionally large training dataset (**BD-L** in Table 1), which includes diagnoses from a large set of clinicians (63 unique clinician codes). This allowed the DL model to "learn" and distill from a variety of clinical assessments. In addition, the DL model's continuous output allows for a calibration against accepted gold standards. Although the distribution of etiologies did differ between training and validation cohorts, performance on the latter remained high, which was sampled directly from clinical distributions. Future work should measure any confounding effect of etiology and/or the presence of liver fibrosis. As well, further investigations should focus on whether DL algorithms like ours can flag patients with NASH from the larger NAFLD population, which would address a universal limitation with current non-invasive steatosis assessments[10].

Importantly, we compare the DL algorithm's performance head-to-head with FibroScan, with is arguably the leading non-invasive steatosis assessment tool today. This was possible because it is CGMH's policy to perform a free FibroScan (since 2016) for patients who receive a liver biopsy at the out-patient department[23]. Comparing cases that have both histology and FibroScan data available, the DL algorithm's "All View Groups" score reported higher AUCROCs (0.91-0.97 vs. 0.80-0.92) and accuracies (77-92% vs. 62-68%) than FibroScan (Table 3, FibroScan comparison study). Statistically significant improvements were achieved for all levels on the **HP-U** cohort, and only for =severe on the **HP-T** cohort. This is encouraging because, unlike FibroScan, our DL algorithm can be applied to many different liver viewpoints and does not require additional equipment outside of an US scanner. Our DL algorithm enjoys similar advantages in flexibility over other quantitative US techniques, such as Canon's attenuation imaging (ATI) technology[42], or backscatter coefficient[20,43], which examines the attenuation, brightness or echogenicity change within a small region of interest (ROI) within the liver parenchyma. By limiting itself to a single ROI, quantitative US might miss other useful information for liver steatosis diagnosis, for example, liver/kidney contrast and the loss of the main portal vein wall. Quantitative US can also not be applied to patients where the specific ROI is unable to be imaged and is often limited to a specific scanner, *e.g.*, ATI only works with Canon scanners.

The DL algorithm can be trained with large-scale retrospective datasets, which need neither costly machines nor annotation of regions of interest. Furthermore, in the inference stage, our algorithm does not require images from all hepatic views, *e.g.*, the absence of the right hepatic lobe images is acceptable, as the performance of our algorithm with left hepatic lobe images is just as good. When using "All View Groups", diagnostic AUCROCs did not improve. However, the improved cross-scanner agreement (see Table 2) suggests using "All View Groups" may provide more reliability. Further investigation is needed. Given the ubiquity of 2D US for screening and diagnostic purposes, the high performance of our DL algorithm suggests that such tools have the potential to enable a quantitative assessment of steatosis that is more universally available than current alternatives. Follow-up technical research should investigate additional capabilities, such as producing a reliably confidence estimate for the steatosis score, which is a critical feature in clinically applied DL[44].

## *Comparison to State of the Art*

The use of machine learning technologies for non-invasive liver steatosis assessment has received attention for a number of years[14–22]. This work represents a significant step forward. Like this work, many prior solutions are built upon recent DL techniques[14,15,17,19–22], which Supplementary Table 7 summarizes. Apart from Gummadi *et al.*[17], all other works only test on a single scanner. Moreover, unlike other works, our evaluation includes more than one etiology: HBV, HCV, and non-hepatitis B/hepatitis C virus (NBNC) liver diseases. In terms of training, we use so-called "big-data", which includes 13 known different US scanner models and over 228K images and 19K US studies. This dwarfs the training set size of other works and should contribute to better model generalizability. Another important point is that most prior studies use one specific viewpoint[14,15,19] or manually defined area of interest[17,20,21] (Supplementary Table 7), which can restrict their applicability. Like Byra *et al.*, we investigated performance across different views[22]. However, we investigate the impact of different liver viewpoints in both hepatic lobes. This advantage is most evident for patients with one resected hepatic lobe or a lobe whose space is occupied by a lesion. Furthermore, a distinct aspect of our work is our reliability analysis: (1) we measure repeatability across different view groups and across different numbers of images per view group and (2) we measure agreement across three different scanners. Quantifying reliability is a critical development for acceptance as a diagnostic tool. To our knowledge we are the first to conduct such an investigation for a DL-based steatosis assessment. Finally, we are the first to compare directly against a leading non-invasive alternative: FibroScan. As such, we are the first to show that a DL algorithm can perform well compared to accepted non-invasive alternatives.

## *Limitations*

Our study has several limitations. First, we primarily use histology as the gold standard for assessing liver steatosis, which can suffer from sampling errors, processing variabilities, and intra and inter-observer variability[28,45]. A histopathological gold standard also introduces patient selection bias. For example, in CGMH histology is not required to initiate therapy in patients with HCV, whereas it is often required for patients with HBV or NASH. Second, for assessing diagnostic performance we used retrospective data. To impose a degree of standardization, we required there to be at least 3 images in a view group and $\geqq 10$ images in the whole US study. Nonetheless, more controlled experimental settings may allow for more precise comparisons and follow-up prospective studies should implement an acquisition protocol to assess diagnostic performance in varied settings. In particular, the relatively poorer repeatability of the LKC view group should be investigated, and, if necessary, adjustments to the protocol should be made. Third, the scanner quality and model does seem to impact the DL model. Future work should better measure this impact, focusing on premium scanners not seen in the training data, which would better characterize generalizability. Relatedly, the data collected in this study were all acquired from the CGMH institution with moderately sized validation cohorts, so any conclusions must be interpreted cautiously. Measuring performance across multiple centers, ideally using a prospective data collection protocol with larger evaluation cohorts, remains an important aspect of future work.

**Total count 5974**

# Supplementary Material

## *Image Preprocessing*

As shown in Supplementary Figure 4, all US images were preprocessed to remove any image regions outside of the actual scan area and to also detect and split images depicting dual US beams. The liver ultrasound image preprocessing pipeline includes 3 steps: image deidentification, background removal, and dual image detection. In the first step, ultrasound images were converted from the DICOM files to PNG files and cropped slightly, to remove protected health information in the DICOM headers and on the boundary of the images. Then in the second step, the most frequent pixel intensity value less than 50 was calculated to identify the background for each ultrasound image, which was then removed. To further crop out the background, for each image after filtering, the largest connected component (LCC) was calculated, and the image was cropped by the smallest square which can hold the LCC. By the end of this step, only the area within the ultrasound region was kept for each image. It is common to see that two ultrasound beams are combined in one saved image, so in the third step, we detect whether dual beams exist in one file. The image was first filtered by the Canny edge filter[1], so only edges were kept, and then a Hough filter[2] was applied to detect the top 8 line segments in the edge map in order to find the borders of the US beams. The intersections between the lines were then calculated, and if an intersection was found that lied near the horizontal center of the image, the image was considered a dual-beam image. This process can be somewhat noisy. However, in a US study there are typically many images of the same type. Therefore, we perform dual-beam detection for each image individually, then we aggregate the results across all images in a study using majority voting. If the majority of images were found to have a dual beam, we split all images in the study using the average intersection location. In Figure 5, an example of a dual-image file is presented, and the intersection point (yellow point in Step 3.b) was used to split the image. For all evaluation datasets, *i.e.*, **HP-U**, **TM**, and **HP-T** datasets, all images were manually verified as being preprocessed correctly.

## *Image selection*

We are interested in investigating performance and reliability across viewpoints. Thus, for all our datasets, we only included US images from the viewpoints shown in Figure 1, which can be labelled as a. left lobe longitudinal, b. left lobe transverse, c. right lobe intercostal, d. lower right lobe intercostal (depicting liver/kidney contrast), e. subcostal depicting liver/kidney contrast, and f. subcostal with hepatic veins views. For the prospective **TM** dataset, we aimed to acquire two US images for each of the six viewpoints of Figure 1, except for the right lobe intercostal viewpoint, where we aimed to acquire four. Occasionally conditions did not allow us to collect certain viewpoints. For **HP-U,** and **HP-T**, we only included studies that had >=10 images of any of the studied viewpoints. As shown in Figure 1, we categorized these six viewpoints into four *view groups*: left liver lobe (LLL), right liver lobe (RLL), liver/kidney contrast (LKC), and subcostal (SC).

Categorizing the view for each image is not necessary for the developmental datasets (**BD-L** and **BD-V)**, as the DL algorithm just trains on each image independently without considering the view. However, even though the specific view for each image need not be categorized, ideally the training set only includes images from the four view groups. Because the **BD-L** and **BD-V** big-data datasets were extracted directly from the CGMH PACS, their US studies may contain images unsuited for liver steatosis analysis, *e.g.*, images of organs other than the liver, liver viewpoints other than those of Figure 1, poor quality images, and even non-US images. So that these non-qualifying images did not impact the training of our DL model, we applied an additional filtering step to remove as many of these images as possible. Given the scale of data, it was not feasible to perform this filtering manually. Instead, we performed this semi-automatically by training a binary DL classifier, using the PyTorch library with hyper-parameters listed in Supplementary Table 4. We first randomly selected 44 US studies (696 images) from **BD-L**, and manually identified the corresponding US images as "qualifying", *i.e.*, belonging to one of the liver viewpoints of Figure 1, or "non-qualifying". We also supplemented the positive training examples using the images within the **HP-U** and **TM** datasets. We then measured the sensitivity and specificity of the trained binary classifier using a mini-validation dataset of 175 images from **BD-L** and chose the operating point corresponding to 95% specificity. *Note, this filtering process was only used to clean the big-data cohorts and was not used for any of the evaluation datasets.*

### *Training Steatosis Assessment DL Algorithm*

Using the images from **BD-L**, we trained a ResNet-18[3] DL classifier using the 2D US diagnoses extracted from the CGMH records. The US diagnoses are ordinal labels ranging from 0 to 3 corresponding to None; Mild; Moderate; and Severe steatosis[4]. Consequently, the learning task is an ordinal regression problem. We treat each image independently in training and follow the well-known binary decomposition approach to ordinal classification of Frank and Hall[5]. As shown in Figure 2, instead of directly regressing the images to a numeric scale or training a four-class classifier, we decompose the problem into three binary classification tasks: estimating the probability the image represents >= mild, >= moderate, or = severe steatosis. Practically, this means that a three-output classification head is used on top of the ResNet-18 backbone. Under this scheme, the scalar labels for None, Mild, Moderate and Severe would be, respectively, converted to (0,0,0), (1,0,0), (1,1,0), and (1,1,1) multi-label vectors. Training is then conducted using standard cross-entropy loss. After training, a simple transformation produces a *continuous* score[6] for each image that ranges from 0 to 1, with higher scores corresponding to more severe steatosis. For a single image, if the model confidences in the Frank and Hall labels are denoted $\hat{y}_i$, where $i$ indexes whether the label is for >=mild, >=moderate, or =severe, then the following formulation produces a severity assessment $\in [0,1]$:

$$\hat{p} = \sum_i \hat{y}_i / 3.0,$$

where $\hat{p}$ represents the image-wise confidence. As Figure 2 indicates, during inference, after feeding the model individual images to obtain image-wise scores, we then take the mean of image-wise scores across each view group to produce a single score for each view group. Additionally, we can also produce an "All View Groups" score, which is the mean score across all view groups in the study.

The hyper-parameters were selected to optimize our algorithm's performance on **BD-V**. Including the convolutional neural network architecture and model optimizer, other hyper-parameters that we tuned include training epochs, initial learning rate, L2 regularization weight, image size and batch size. The details of these hyper-parameters are specified in Supplementary Table 5. We also applied an aggressive augmentation scheme to increase the variability in the image distribution presented to the network. These include additive Gaussian noise, brightness and contrast jittering, and random rotations. Each augmentation was applied on-the-fly to an image with a 50% probability. We also executed an aggressive cropping augmentation. Finally, all images were resampled to 256x256 pixels before being inputted into the deep neural network.

## More Details on the Reliability Study

**Repeatability Study (Experiment 1)**

We used **TM** and **HP-U** to assess how many images are needed per view group to achieve repeatability. Note, for the **TM** dataset we randomly selected only one US study for each patient to avoid sampling the same patient more than once. Typically, to calculate repeatability one simply acquires repeated measurements and performs an accepted repeatability metric. However, in our case each measurement can itself consist of the mean measurement across several image-wise scores. For example, if we are interested in the repeatability when averaging the score across three images to calculate a view group score, then two view-group measurements would require acquiring six images. This is an onerous data collection requirement. Instead of doing this, we simply first calculate the within-subject standard deviation, $s$, of the image-wise scores. We do this for each US study, which gives us a set of $s$ values across different mean severity measurements. If we are then taking the mean across $k$ images to obtain a view-group score, the resulting within-subject standard deviation is simply $\sqrt{1/k} \times s$. Finally, the within-subject standard deviation of differences between repeated measurements can be estimated as $s_k = \sqrt{2/k} \times s$. The advantage of such an approach is that the within-subject standard deviation can be calculated for any $k$ without requiring the collection of more images.

As advocated by Bland and Altman[7,8], these within subject standard deviations were then graphed across different view-group steatosis scores. Typically the repeatability coefficient (RC) could then be calculated using a mean $s_k$ value across all US studies[8]: the difference between two repeated measurements should be within the RC value for 95% of the US studies. However, because $s_k$ is not uniform (typically greater variability at moderate steatosis levels), a uniform RC is not appropriate[8]. Instead, we modelled the heteroskedasticity by regressing the within-subject standard deviation on mean severity scores[8,9] using a cubic regression. We chose a cubic regression because there is a skew in the distribution of $s_k$ values (see Supplementary Figure 1). We then used the worst-case RC value (max RC) as a summary statistic, with 95% confidence intervals computed using percentile bootstrap (1000 bootstrap samples)[10]. We conducted this for k = {1,2,3,4} and for every view group.

**Cross-Scanner Agreement (Experiment 2)**
We evaluated agreement across scanners using the *TM* dataset, which consists of multiple studies taken on the same day of the same patient. A Bland-Altman analysis[29,30] was performed for assessing cross-scanner agreement. This was simpler than what was done for repeatability, since for a chosen view group we just computed the mean score across all available images in a study. However, based on the repeatability measurements of Experiment 1, we only included view group scores with >=3 images. Thus, for two studies of the same patient across two different scanners, there are only two observations to compare. We calculated the bias and LOAs, where the latter are the limits by which 95% of the disagreements fall under[30]. To deal with the same heteroskedasticity faced by the repeatability experiment, we regressed non-uniform limits of agreement (LOAs)[30] and used the maximum upper LOAs and minimum lower LOAs as summary statistics. 95% confidence intervals were computed using the same bootstrap approach as in Experiment 1.

# Tables and Figures:

**Supplementary Table 1a.** Demographic Features of Each Cohort

|  | BD-L | BD-V | TM | HP-U | HP-T |
|---|---|---|---|---|---|
| Number of Patients | 2899 | 411 | 246 | 147 | 112 |
| Number of Studies | 17149 | 2364 | 733 | 147 | 112 |
| Number of Images | 200654 | 27421 | 9215 | 1647 | 1996 |
|  |  |  |  |  |  |
| Mean Age at Scan | 56.5 | 56.9 | 56.6 | 49.1 | 50.0 |
| Male, n (%) | 1752 (60.4) | 248 (60.3) | 157 (63.8) | 93 (63.3) | 66 (58.9) |
| Female, n (%) | 1147 (39.6) | 163 (39.7) | 89 (36.2) | 54 (36.7) | 46 (41.1) |
|  |  |  |  |  |  |
| NBNC, n (%) | 353 (12.2) | 51 (12.4) | 56 (22.8) | 103 (70.1) | 63 (56.2) |
| HBV, n (%) | 1050 (36.2) | 145 (35.3) | 125 (50.8) | 35 (23.8) | 46 (41.1) |
| HCV, n (%) | 1322 (45.6) | 190 (46.2) | 65 (26.4) | 9 (6.1) | 3 (2.7) |
| Others/Unknown, n (%) | 174 (6.0) | 25 (6.1) | 0 (0.0) | 0 (0.0) | 0 (0.0) |
|  |  |  |  |  |  |
| *Steatosis Grade* |  |  |  |  |  |
| *US* grade 0, n (%) | 2529 (87.2) | 352 (85.6) | N/A | N/A | N/A |
| *US* grade 1, n (%) | 314 (10.8) | 50 (12.2) | N/A | N/A | N/A |
| *US* grade 2, n (%) | 50 (1.7) | 8 (1.9) | N/A | N/A | N/A |
| *US* grade 3, n (%) | 6 (0.3) | 1 (0.3) | N/A | N/A | N/A |

**Supplementary Table 1b.** Additional Clinicopathologic Features of *HP-U* and *HP-T*

|  | HP-U | | | HP-T | | |
|---|---|---|---|---|---|---|
|  | **NBNC** | **HBV** | **HCV** | **NBNC** | **HBV** | **HCV** |
| Number of Patients | 103 | 35 | 9 | 63 | 46 | 3 |
| Mean Age at Scan | 47.5 | 51.8 | 56.6 | 48.8 | 52.2 | 43.2 |
| Male, n (%) | 71 (68.9) | 18 (51.4) | 4 (44.4) | 28 (44.4) | 36 (78.3) | 2 (66.7) |
| Female, n (%) | 32 (31.1) | 17 (48.6) | 5 (55.6) | 35 (55.6) | 10 (21.7) | 1 (33.3) |
| Mean BMI | 27.5 | 25.4 | 27.1 | 25.7 | 25.8 | 27.5 |
| Mean AST U/L | 64.9 | 64.1 | 58.0 | 115.4 | 87.0 | 71.7 |
| Mean ALT U/L | 110.2 | 92.7 | 76.4 | 213.4 | 151.8 | 128.0 |
| Mean PLT $10^3/mm^3$ | 246.9 | 201.4 | 207.3 | 248.8 | 186.2 | 179.7 |
| ***Steatosis Grade*** | | | | | | |
| grade 0, n (%) | 10 (9.7) | 11 (31.4) | 3 (33.3) | 22 (34.9) | 21 (45.7) | 1 (33.3) |
| grade 1, n (%) | 18 (17.5) | 14 (40.0) | 4 (44.4) | 13 (20.6) | 15 (32.6) | 1 (33.3) |
| grade 2, n (%) | 31 (30.1) | 4 (11.4) | 0 (0.0) | 6 (9.6) | 8 (17.4) | 0 (0.0) |
| grade 3, n (%) | 44 (42.7) | 6 (17.1) | 2 (22.2) | 22 (34.9) | 2 (4.3) | 1 (33.3) |
| ***Fibrosis Grade*** | | | | | | |
| grade 0, n (%) | 20 (19.4) | 2 (5.7) | 1 (11.1) | | | |
| grade 1, n (%) | 56 (54.4) | 9 (25.7) | 3 (33.3) | | | |
| grade 2, n (%) | 5 (4.9) | 11 (31.4) | 0 (0.0) | | | |
| grade 3, n (%) | 17 (16.5) | 9 (25.7) | 4 (44.4) | | | |
| grade 4, n (%) | 5 (4.9) | 4 (11.4) | 1 (11.1) | | | |

***AST***: aspartate aminotransferase; ***ALT***: alanine aminotransferase; ***HBV***: hepatitis B; ***HCV***: hepatitis C; ***NBNB***: non-HBV, non-HCV and excluded other liver diseases (E.g. alcoholic, autoimmune, etc); ***PLT***: platelet

**Supplementary Table 2.** Scanner brands, number of studies, and time ranges (if information is available in de-identified DICOM headers)

| Scanner Brand | BD-L, BD-V Studies | BD-L, BD-V Time Range | HP-U Studies | HP-U Time Range | HP-T Studies | HP-T Time Range |
|---|---|---|---|---|---|---|
| ATL: HDI 5000 | 2865 | 1/3/2011 – 4/13/2015 | -- | -- | 16 | 3/18/2011 – 9/2/2014 |
| GE Healthcare: LOGIQ E9 | 2 | 11/11/2014– 11/14/2014 | -- | -- | -- | -- |
| GE Healthcare: LOGIQ S8 | 19 | 8/29/2012 – 9/5/2012 | -- | -- | -- | -- |
| Aloka Medical,Ltd.: SSD 5500 | 4273 | 1/3/2011 – 10/17/2014 | -- | -- | 2 | 5/2/2012 – 3/19/2013 |
| Hitachi Medical Corporation: HI VISION Avius | 16 | 8/27/2012 – 8/31/2012 | -- | -- | -- | -- |
| Hitachi Medical Corporation: HI VISION Preirus | 20 | 7/18/2012 – 9/25/2018 | -- | -- | -- | -- |
| Philips Medical Systems: EPIQ 7G | 2 | 11/21/2014 – 7/24/2018 | -- | -- | -- | -- |
| Philips Medical Systems: HD15 | 4 | 11/17/2014 – 11/20/2014 | -- | -- | -- | -- |
| Philips Medical Systems: iU22 | 8827 | 1/3/2011 – 9/28/2018 | 7 | 11/27/2014 – 6/19/2019 | 12 | 9/9/2011 –9/4/2020 |
| Siemens: S2000 | 193 | 1/6/2011 – 9/28/2018 | 117 | 7/12/2012 – 9/26/2019 | 78 | 8/14/2012 – 1/29/2021 |
| SuperSonic Imagine SA: Aixplorer | 72 | 5/14/2012 – 7/24/2012 | -- | -- | -- | -- |
| Toshiba MEC US: TUS-A300 | 3145 | 11/20/2014 – 9/28/2018 | 23 | 7/14/2015 – 7/2/2019 | 4 | 6/9/2020 – 12/16/2020 |
| Toshiba MEC: Xario | 26 | 8/24/2012 – 8/31/2012 | -- | -- | -- | -- |
| Unknown * | 49 | 1/4/2011-9/28/2018 | -- | -- | -- | -- |

\* Unknown: Toshiba SSA-370A or Toshiba SSA-700A, the exact model used was not recorded.

**Supplementary Table 3.** Performance Statistics for All Experiments Described in This Article. All experiments evaluated the same model, trained on the *BD-L* dataset.

| ID | Experiment description | Result statistics |
|---|---|---|
| 1 | Estimate repeatability across view groups and different numbers of images per view group using two *TM* and *HP-U* cohorts | Max repeatability coefficient (RC), RC graphs |
| 2 | Estimate consistency across scanners and view groups using *TM* cohort | Bias, upper and lower limits of agreement, Bland-Altman graphs, % Agreement |
| 3a | Estimate diagnostic performance across views using histology proven cohort *HP-U* | AUCROC (fatty % >=5%; >=33%; and >=66%), ROC Curves, Accuracy |
| 3b | Compare diagnostic performance of DL model to FibroScan using studies with associated FibroScan scores from the *HP-U* cohort | AUCROC (fatty % >=5%; >=33%; and >=66%), ROC curves, Accuracy |
| 4a | Estimate diagnostic performance across views using histology proven cohort *HP-T* | AUCROC (fatty % >=5%; >=33%; and >=66%), ROC curves, Accuracy |
| 4b | Compare diagnostic performance of DL model to FibroScan using studies with associated FibroScan scores from the *HP-T* cohort | AUCROC (fatty % >=5%; >=33%; and >=66%), ROC curves, Accuracy |

**Supplementary Table 4.** Description and values of all hyperparameters and properties of the image quality binary classifier. This DL model was used to automatically filter out non-qualifying images from the ***BD-L*** and ***BD-V*** dataset.

| Hyperparameter | Description | Value |
| --- | --- | --- |
| Network architecture | Deep neural network layout | ResNet-18 |
| Image size | Size of image as the network input (in pixel) | 256×256 |
| Maximum Epochs | Maximum number each image is shown to the network during training | 100 |
| Graphics Processing Unit | Graphics processing unit hardware | NVIDIA Titan V |
| Initial Learning Rate | Network learning rate during training | 0.0001 |
| L2 Regularization | Weight decay (L2 penalty) | 0.0005 |
| Batch Size | Number of images processed in parallel | 16 |
| Solver | Optimizer to update weights and biases | SGD |

**Supplementary Table 5.** Description and values of all hyperparameters and properties of the deep learning workflow for steatosis severity assessment.

| Hyperparameter | Description | Value |
| --- | --- | --- |
| Network architecture | Deep neural network layout | ResNet-18 |
| Image size | Size of image as the network input (in pixel) | 256×256 |
| Maximum Epochs | Maximum number each image is shown to the network during training | 120 |
| Graphics Processing Unit | Graphics processing unit hardware | NVIDIA Titan V |
| Initial Learning Rate | Network learning rate during training | 0.0005 |
| L2 Regularization | Weight decay (L2 penalty) | 0.0001 |
| Batch Size | Number of images processed in parallel | 32 |
| Solver | Optimizer to update weights and biases | SGD |
| Gaussian Noise | Standard deviation upper bound | 0.01 |
| Color Jittering | Brightness/Contrast change upper bound | 0.2 |
| Rotation | Affine transformation rotation upper bound | 10 Degrees |
| Scaling | Affine transformation ratio bound | [0.9, 1,1] |
| Augmentation Possibility | The possibility to apply each augmentation technique to a single image | 50% |

**Supplementary Table 6.** The max repeatability coefficient (RC) is tabulated across different view groups for the *TM* and *HP-U* datasets. "All Views" represents the DL-based score when taking the average across all view-group scores. Parentheses enclose bootstrapped 95% confidence intervals.

| View | 1 Image | 2 Images | 3 Images | 4 images |
|---|---|---|---|---|
| LLL | 0.46 (0.42, 0.51) | 0.33 (0.30, 0.36) | 0.27 (0.24, 0.29) | 0.23 (0.21, 0.26) |
| RLL | 0.37 (0.34, 0.40) | 0.26 (0.24, 0.28) | 0.21 (0.20, 0.23) | 0.18 (0.17, 0.20) |
| LKC | 0.53 (0.47, 0.58) | 0.37 (0.33, 0.41) | 0.30 (0.27, 0.34) | 0.26 (0.24, 0.29) |
| SC | 0.46 (0.42, 0.50) | 0.32 (0.30, 0.36) | 0.27 (0.24, 0.29) | 0.23 (0.21, 0.26) |

**Supplementary Table 7.** Literature review of works applying deep learning techniques for assessing hepatic steatosis using 2D US images. To be included, the works must be using deep learning models and only the deep learning results are highlighted here.

| Reference | Byra et al.[22] | Chen et al.[20] | Cao et al.[21] | Han et al.[19] | Gummadi et al.[17] | Byra et al.[14] | Biswas et al.[15] | Ours |
|---|---|---|---|---|---|---|---|---|
| Reference in Main Body | 22 | 20 | 21 | 19 | 17 | 14 | 15 | |
| Publication Year | 2021 | 2020 | 2020 | 2020 | 2020 | 2018 | 2018 | |
| Evaluation Studies (case/control) | 135‡ | 41 | 240 (138/106) | 204 (140/64) | Unclear patient or study-wise split | 55 (38/17)‡ | 63 (36/27) ‡ | 147+112 |
| Training Studies | Leave-one-out cross validation | 164 | ? | | Unclear patient or study-wise split | Leave-one-out cross validation | Ten-fold cross validation, unclear if split across patients | 19,513 |
| Etiology | NBNC | NBNC | NAFLD | NAFLD | NAFLD/NASH | Severely obese | NAFLD | HBV/HCV/NBNC |
| US Scanner | Siemens S3000 | Terason M3000 | Mindray Resona 7 | Siemens S2000 | 6 models | GE | US Scanner | 13 models (training); 5 models (evaluation) |
| Image type | Grayscale | Grayscale | Grayscale | RF data | Grayscale | Grayscale | Grayscale | Grayscale |
| Evaluation Images/Case | 4 | 5 | ? | 10 | ~5 | 10 | ? | >= 10 |
| Total Evaluation Images | 540 | 205 | ? | 2040 | 78 | 550‡ | ? | 1647+1996 |
| Total Training Images | Leave-one-out cross validation | 820 | 852 | | 725 | Leave-one-out cross validation | Ten-fold cross validation, unclear if split across patients | 228075 |
| Area of interest | Cropped 224x224 pixel image | Manual 3.5*3.5 cm ROI | Manual 224*224 pixels ROI | 256 RF signals | Manual Crop | 434×636 pixel Image | Auto-cropped 128*128 Image | Auto-cropped 256*256 pixel Image |
| Gold Standard | MRI PDFF | Histology | 2D-US | MRI PDFF | Histology/MRI PDFF/2D-US & Patient History | Histology | Normal control | Histology |
| View | RLL and LKC | RLL | All Views | RLL | Unspecified | RLL/Kidney | RLL | All Views |
| Machine Learning Model | ResNet-50 | VGG-16 | Custom CNN | CNN | Unspecified CNN | Pretrained Inception-ResNetv2 CNN+SVM | SVM/ELM/CNN | ResNet-18 |
| Results | | | | | | | | |
| AUCROC | | 0.71† | 0.933* | | | | | 0.85-0.95 |

| | | | | | | | |
|---|---|---|---|---|---|---|---|
| AUCROC (mild) | | 0.75† | 0.692* | | | | 0.91-0.92 |
| AUCROC (moderate) | | | | | | | |
| AUCROC (severe) | | 0.88† | 0.958* | | | | 0.87-0.93 |
| AUCROC Binary | 0.86-0.91 | | 0.98 | 0.98 | 89% Se. & 95% Sp. | 0.98 | 1.0 |

\* Based on 2D-US diagnosis;  † Separate data splits for each cut-off;  ‡ Cross validated

Abbreviations: AUCROC: area under the curve of receiver operating characteristic; CNN: convolutional neural network; HBV: hepatitis virus B; HCV: hepatitis virus C; MRI PDFF magnetic resonance imaging derived proton density fat fraction; NBNC: non-hepatitis B/non-hepatitis C; RF: Radiofrequency;  RLL: right liver lobe;  LKC: liver kidney contrast; SVM: support vector machine

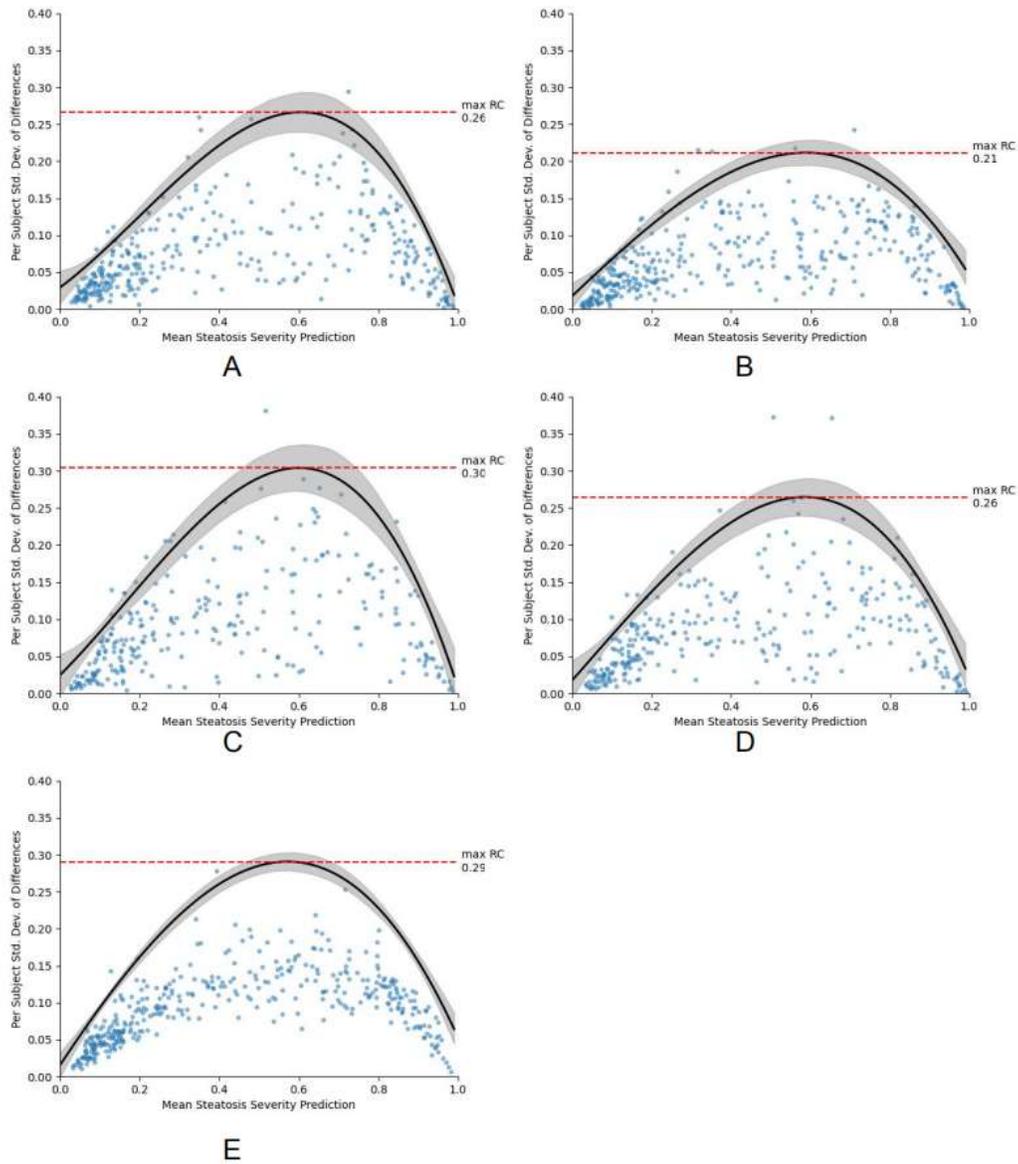

**Supplementary Figure 1**. Repeatability coefficient (RC) plot across different 2D US viewpoints. (A) to (E) represents LLL, RLL, LKC, SC, and "All View Groups" respectively. Repeatability is measured when taking the mean score across three images per view group. "All View Groups" represents the score after taking the mean each resulting viewpoint score to create one score for each study.

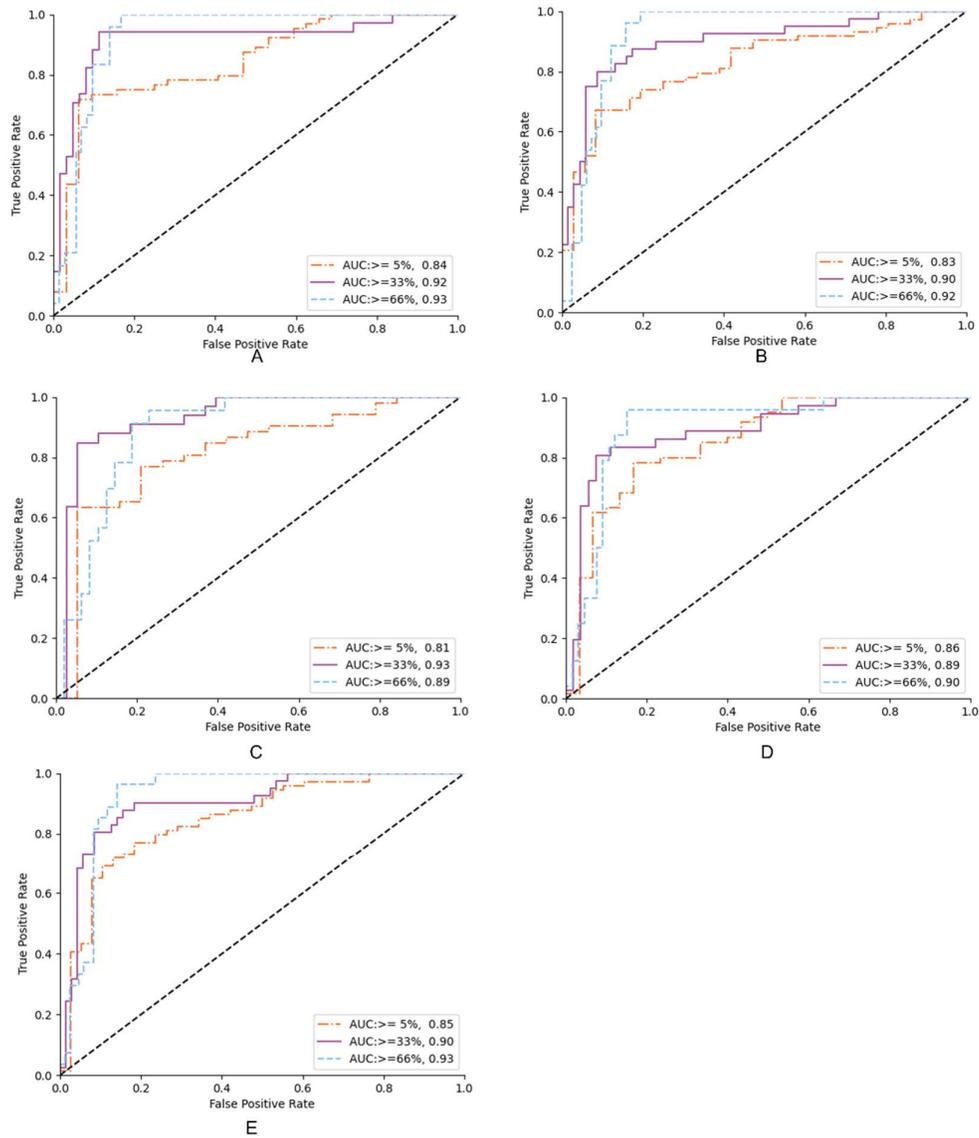

**Supplementary Figure 2:** ROC analysis of *HP-T* (Individual view group setting). (A) to (E) shows ROC curves of the DL model for diagnosing hepatic steatosis grades on HP-T with LLL, RLL, LKC, SC, and "All View Groups", respectively.

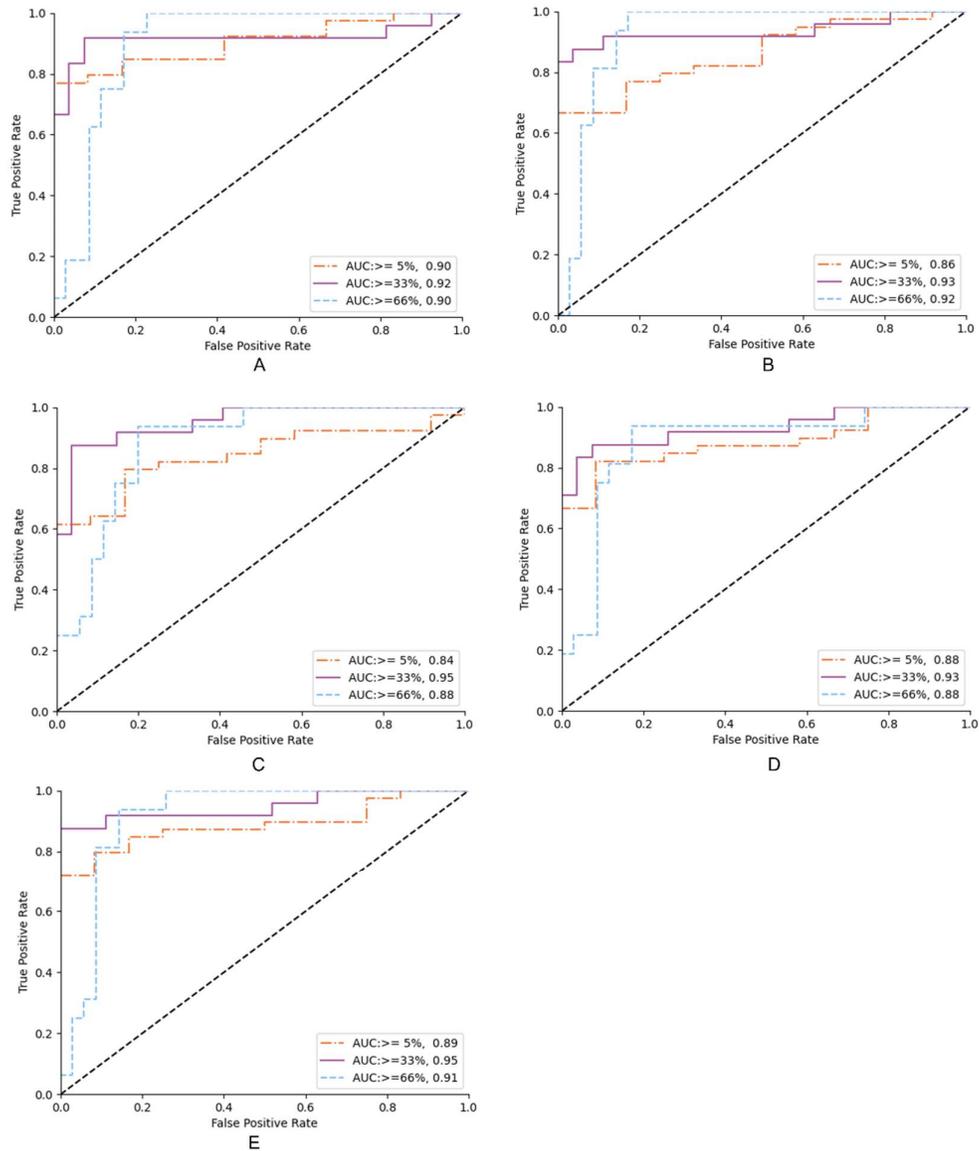

**Supplementary Figure 3:** ROC analysis of *HP-T* (Complete view group setting). (A) to (E) shows ROC curves of the DL model for diagnosing hepatic steatosis grades on HP-T with LLL, RLL, LKC, SC, and "All View Groups", respectively.

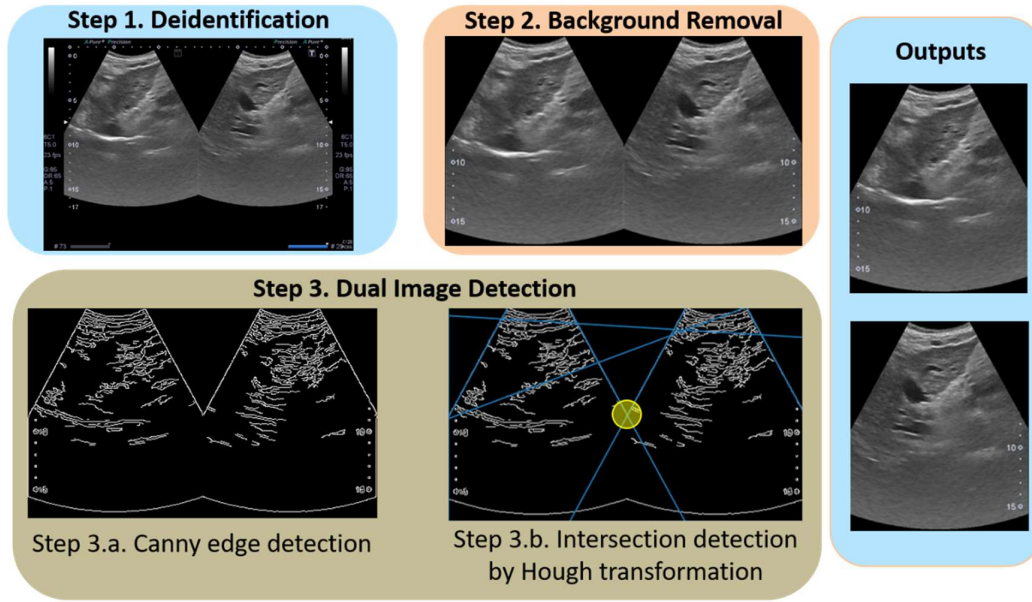

**Supplementary Figure 4:** Liver ultrasound image preprocessing pipeline includes 3 steps: image deidentification, background removal, and dual image detection. In "Step 3.b", the figure is showing the top 8 lines detected by the Hough transform (in blue, two lines are along the boundaries and might not be seen), and the detected intersection point (in yellow).